\documentclass[epj, twocolumn]{webofc}

\usepackage[varg]{txfonts} 
\usepackage{graphicx}
\wocname{epj}
\woctitle{Seismology of the Sun and the Distant Stars 2016}

\begin{document}
\title{Seismic diagnosis from gravity modes strongly affected by rotation}
\author{\firstname{Vincent} \lastname{Prat}\inst{1}\fnsep\thanks{\email{vincent.prat@cea.fr}} \and \firstname{St\'ephane} \lastname{Mathis}\inst{1} \and \firstname{Fran\c cois} \lastname{Ligni\`eres}\inst{2,3} \and \firstname{J\'er\^ome} \lastname{Ballot}\inst{2,3} \and \firstname{Pierre-Marie} \lastname{Culpin}\inst{2,3}}
\institute{Laboratoire AIM Paris-Saclay, CEA/DRF - CNRS - Universit\'e Paris Diderot, IRFU/SAp Centre de Saclay, F-91191 Gif-sur-Yvette, France
\and
Universit\'e de Toulouse; UPS-OMP; IRAP; Toulouse, France \and CNRS; IRAP; 14 avenue \'Edouard Belin; F-31400 Toulouse, France}
\abstract{
Most of the information we have about the internal rotation of stars comes from modes that are weakly affected by rotation, for example by using rotational splittings.
In contrast, we present here a method, based on the asymptotic theory of Prat et al. (2016), which allows us to analyse the signature of rotation where its effect is the most important, that is in low-frequency gravity modes that are strongly affected by rotation.
For such modes, we predict two spectral patterns that could be confronted to observed spectra and those computed using fully two-dimensional oscillation codes.
} 
\maketitle
\section{Introduction}
\label{intro}

Stellar evolution plays a central role in astrophysics by linking observable quantities, such as effective temperature or surface gravity, to fundamental parameters of stars that can be used to constrain other objects such as galaxies or planets.
In this context, angular momentum and chemical transport processes are crucial for stellar evolution, since they modify the structure of stars (\cite{Maeder}).
For instance, rotation has three important effects: (i) it directly modifies the structure of stars because of the flattening due to the centrifugal acceleration, (ii) it induces many hydrodynamical instabilities, such as the shear instability, that create turbulent transport, and (iii) it drives large-scale flows such as meridional circulation (\cite{Zahn, MaederZahn, MathisZahn, EspinosaRieutord}).

Surface rotation can be investigated by different methods, but only asteroseismology gives access to internal properties of stars such as deep differential rotation and chemical stratification.
For slowly rotating stars, perturbative methods can be used to estimate internal rotation by measuring the splittings between modes of different azimuthal orders.
This has been done for the Sun (e.g. \cite{Schou,Garcia07}), subgiant and red giant stars (e.g. \cite{Deheuvels15}), solar-type stars (\cite{Benomar}) and a few intermediate-mass (e.g. \cite{Saio15}) and massive stars (\cite{Triana}).
For rapidly rotating stars, because the problem of finding eigenmodes becomes intrinsically two-dimensional, especially for gravity modes, such methods fail (see e.g. \cite{Ballot10}) and new diagnostic tools need to be developed.
The traditional approximation, which neglects the latitudinal component of the rotation vector, allows for variable separation for all rotations (see e.g. \cite{Townsend03}), thus providing such a tool, but its relevance for seismology is uncertain (see e.g. \cite{Gerkema}).

One way to go further is to use asymptotic theories that rely on some assumptions to gain physical insight.
For example, it is possible to describe waves as rays in a way similar to geometrical optics by assuming that the wavelength is much smaller than the characteristic length of the variations of the background.
These rays can be used to explore the properties of modes more efficiently than full numerical computations.
Such an asymptotic theory has been built for acoustic waves (\cite{LG09}) and has led to the prediction of regularities in corresponding spectra of fast-rotating stars (\cite{Pasek12}).
For gravito-inertial waves, a first study (\cite{PratLB}) has allowed us to constrain the structure of the phase space.
In particular, we have shown that it is dominated by nearly integrable structures at low frequencies.
This suggests the existence of a nearby integrable system that could be used to derive new diagnoses.

The aim of this article is to present a new low-frequency approximation (Sect.~\ref{sec:lfa}) that leads to an integrable and separable system.
Then, we use semi-classical quantisation techniques to determine mode frequencies and resulting spectral patterns (Sect.~\ref{sec:diagn}).
Finally, we conclude and discuss the prospects of this work in Sect.~\ref{conclu}.

\section{Low-frequency approximation}
\label{sec:lfa}

In \cite{PratLB}, we applied the Jeffreys-Wentzel-Kramers-Brillouin (JWKB) small-wavelength approximation to the linearised equations describing the dynamics of gravito-inertial waves, and we obtained the general eikonal equation (i.e. local dispersion relation)
\begin{equation}
    \omega^2 = \frac{f^2k_z^2+N_0^2\left(k_\perp^2+k_\phi^2\right)+f^2\cos^2\Theta k_{\rm c}^2}{k^2+k_{\rm c}^2},
\end{equation}
where $\omega$ is the wave pulsation, $f=2\Omega$ with $\Omega$ the rotation rate, $N_0$ is the Brunt-V\"ais\"al\"a frequency, $k_z$, $k_\perp$ and $k_\phi$ are the components of the wave vector $\vec k$ parallel to the rotation axis, orthogonal to the effective gravity (in the meridional plane) and in the azimuthal direction, respectively; $k$ is the norm of the wave vector, $k_{\rm c}$ is a term that accounts for the back-refraction of waves near the surface, and $\Theta$ is the angle between the rotation axis and the effective gravity (equal to the colatitude $\theta$ when centrifugal deformation is neglected).

As shown in \cite{PratLB}, low-frequency gravito-inertial waves are trapped near the equatorial plane, but can propagate near the centre, which is not accounted for by the traditional approximation.
In the limit of very low frequencies ($\omega\ll f$), the trapping implies that
\begin{equation}
    \frac{N_0^2}{N_0^2+f^2}\cos^2\Theta \ll 1.
\end{equation}
It is possible to use this condition to simplify the general eikonal equation in the vicinity of the equatorial plane.
After some manipulations, one can finally write
\begin{equation}
    \omega^2 = \frac{f^2\cos^2\delta\left(k_\beta^2 + k_{\rm c}^2\right) + \left(N_0^2+f^2\right)\frac{k_\delta^2+m^2}{\zeta^2}}{k^2+k_{\rm c}^2},
\end{equation}
where
\begin{equation}
    \cos^2\delta = \frac{N_0^2}{N_0^2+f^2}\cos^2\Theta,\quad\zeta = \frac{r\sqrt{N_0^2+f^2}}{N_0},
\end{equation}
$m$ is the azimuthal order, $\beta$ is the principal coordinate which is equivalent to $r\sin\theta$ near the centre and to $r$ in the envelope, and $k_\beta$ and $k_\delta$ are the natural components of $\vec k$ associated with $\beta$ and $\delta$, respectively.

The next step is to describe the ray dynamics using the Hamiltonian formalism given by
\begin{eqnarray}
    \frac{{\rm d}x_i}{{\rm d}t} &=& \frac{\partial\omega}{\partial k_i},\\
    \frac{{\rm d}k_i}{{\rm d}t} &=& -\frac{\partial\omega}{\partial x_i},
\end{eqnarray}
where $k_i=\partial\Phi/\partial x_i$ is the natural component of the wave vector associated with the coordinate $x_i$.
These equations ensure that the frequency $\omega$ is invariant.
To go further, we make the simplification that $k_{\rm c}$, $N_0$ and $\zeta$ depend only on $\beta$.
This is rigourously valid only at zero rotation, but should be a good approximation in the vicinity of the equatorial plane, at least for moderate rotation rates.
One then can show that the quantity
\begin{equation}
    \chi = \frac{N_0^2+f^2\sin^2\delta}{\zeta^2(k^2+k_{\rm c}^2)}
\end{equation}
is also invariant.
Thus, the system has two invariants, $\omega$ and $\chi$, which means it is integrable.
Furthermore, the ray dynamics can be described in a separable way,
\begin{eqnarray}
    k_\beta^2 + k_{\rm c}^2 &=& \frac{N_0^2+f^2-\omega^2}{\zeta^2\chi},  \label{eq:kbe}  \\
    k_\delta^2              &=& \frac{\omega^2-f^2\cos^2\delta}{\chi}-m^2,   \label{eq:kde}
\end{eqnarray}
in the sense that each component of the wave vector can be expressed as a function of the corresponding coordinate only.

\begin{figure}
    \includegraphics[width=\hsize,clip]{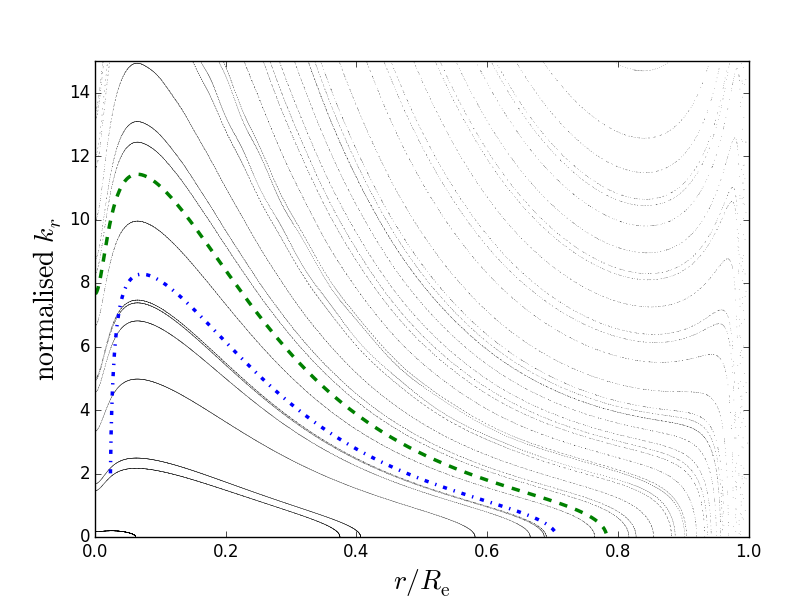}
    \caption{Poincar\'e surface of section at $\omega/f=0.8$ of a polytropic model of star (with a polytropic index of 3) rotating at 38\% of its critical velocity. The blue dash-dotted line and the green dashed line are computed using the traditional approximation and of the low-frequency approximation, respectively.}
    \label{fig:pss}
\end{figure}

At the equator, $k_\beta=k_r$ and $\beta=r$, so Eq.~(\ref{eq:kbe}) can be used to compare the present approximation with the full dynamics.
This is illustrated in Fig.~\ref{fig:pss}, where one can see that the low-frequency approximation reproduces well the structure of a Poincar\'e surface of section (see \cite{PratLB} for explanations) with the full dynamics, whereas the traditional approximation fails near the centre.

\section{Seismic diagnoses}
\label{sec:diagn}

To construct modes from rays, one must satisfy quantisation conditions that ensure that when a ray crosses itself, its phase is the same.
They read
\begin{equation}
    \int_\mathcal{C}\vec k\cdot{\rm d}\vec x = 2\pi\left(p+\frac{\varepsilon}{4}\right),
\end{equation}
where $\mathcal{C}$ is a closed contour, $p$ and $\varepsilon$ are integers, and $\varepsilon$ accounts for phase-shifts due to boundary conditions.
In our case, there are two independent sets of contours, which yield a non-linear system in $\omega$ and $\chi$ to solve: 
\begin{align}
    \int_0^{r_{\rm s}}\sqrt{\frac{N_0^2+f^2-\omega^2}{\zeta^2\chi}-k_{\rm c}^2}\ {\rm d}r                 &= \frac{\pi}{2}\left(\tilde n+\frac14\right),  \label{eq:qb}\\
    \int_{\delta_{\rm c}}^{\frac{\pi}{2}}\sqrt{\frac{\omega^2-f^2\cos^2\delta}{\chi}-m^2}\ {\rm d}\delta  &= \frac{\pi}{2}\left(\tilde\ell+\frac12\right),
\end{align}
where $r_s$ is the stellar radius, $\delta_{\rm c}=\arccos\frac{\sqrt{\omega^2-m^2\chi}}{f}$ the critical value of $\delta$, $\tilde n=n-1$ and $\tilde l = l_\mu-1$ with $n$ the radial order and $l_\mu$ the number of nodes in the latitudinal direction.

This system can be solved numerically, but it is possible to solve it analytically by making further assumptions, such as using the fact that $\omega\ll f$ and neglecting the influence of $k_{\rm c}$ in Eq.~\eqref{eq:qb}.
This leads to
\begin{equation}
    \omega^2 = \frac{(2\tilde\ell+1)f\int_0^{r_{\rm s}}\frac{N_0}{r}{\rm d}r}{\pi\left(\tilde n+\frac14\right)} + m^2\frac{\left(\int_0^{r_{\rm s}}\frac{N_0}{r}{\rm d}r\right)^2}{\pi^2\left(\tilde n+\frac14\right)^2},
\end{equation}
expressed in the co-rotating frame.

For axisymmetric modes ($m=0$), the period spacings between modes of equal $\tilde l$ and consecutive $\tilde n$ are
\begin{equation}
    \Delta\Pi \simeq\frac{\pi^{3/2}}{\sqrt{2\left(\tilde n+\frac34\right)(2\tilde\ell+1)\Omega\int_0^{r_{\rm s}}\frac{N_0}{r}{\rm d}r}}.
\end{equation}
They allow for measurements of the product $\Omega\int_0^{r_{\rm s}}\frac{N_0}{r}{\rm d}r$.
For non-axisymmetric modes, the period spacings in the inertial frame read
\begin{equation}
    \Delta\Pi \simeq \frac{2}{m^2}\sqrt{\frac{\left(\tilde\ell+\frac12\right)\pi\int_0^{r_{\rm s}}\frac{N_0}{r}{\rm d}r}{\Omega^3\left(\tilde n+\frac34\right)^3}},
\end{equation}
and they allow for measurements of $\left(\int_0^{r_{\rm s}}N_0{\rm d}r/r\right)/\Omega^3$.

\section{Conclusion}
\label{conclu}

Both kind of spacings contain information on the internal rotation and on the stratification of the star, but with different scalings laws.
If both are measured, rotation and stratification can be constrained at the same time.
When using stellar models in which stratification is given, only one kind of spacings is needed to constrain rotation.

These spacings could be used to interpret observed spectra of fast rotators exhibiting g modes with frequencies lower than their rotation frequency, such as $\gamma$ Doradus, SPB or Be stars (see studies in the traditional approximation: \cite{VanReeth} for $\gamma$ Doradus stars, and \cite{Moravveji} for a SPB star).

The new approximation on which is based our predicted spectral patterns is similar to the traditional approximation.
However, our approximation is more accurate near the centre, as it allows waves to propagate there, and it is not limited to spherical models.

This work has been done under the assumption of uniform rotation.
The next step is to investigate the case of differentially rotating stars (see \cite{Mathis09, Mirouh}).

\vskip1em
\begin{acknowledgement}
    V.P. and S.M. acknowledge funding by the European Research Council through ERC grant SPIRE 647383.
    The authors acknowledge funding by SpaceInn and PNPS (CNRS/INSU) and by CNES CoRoT/PLATO grant at SAp and IRAP.
\end{acknowledgement}

\bibliography{Prat}

\end{document}